\documentclass[prd,preprint,tightenlines,floatfix,showpacs,preprintnumbers,nofootinbib,eqsecnum]{revtex4}
 \usepackage[dvips,final]{graphicx}
  \usepackage{amssymb}
   \usepackage{amsmath}
    \usepackage{epsfig}
     \usepackage{bm}

\begin{document}
\thispagestyle{empty} \preprint{\hbox{}} \vspace*{-10mm}

\title{Dispersive approach to the axial anomaly \\
and nonrenormalization theorem}

\author{R.~S.~Pasechnik}
\email{rpasech@theor.jinr.ru} \altaffiliation[\\Also at ]{Faculty
of Physics, Moscow State University, 119992 Moscow, Russia}

\author{O.~V.~Teryaev}
\email{teryaev@theor.jinr.ru}

\affiliation{ Bogoliubov Laboratory of Theoretical Physics, JINR,
Dubna 141980,Russia}

\date{\today}

\begin{abstract}
Anomalous triangle graphs for the divergence of the axial-vector
current are studied using the dispersive approach generalized for
the case of higher orders of perturbation theory. The validity of
this procedure is proved up to two-loop level. By direct
calculation in the framework of dispersive approach we have
obtained that the two-loop AVV amplitude is equal to zero.
According to the Vainshtein's theorem the transversal part of the
anomalous triangle is not renormalized in the chiral limit. We
generalize this theorem for the case of finite fermion mass in the
triangle loop.
\end{abstract}

\pacs{12.38.Bx; 11.55.Fv}


\maketitle

\section{Introduction}

\begin{figure}[h]
 \centerline{\includegraphics[width=0.35\textwidth]{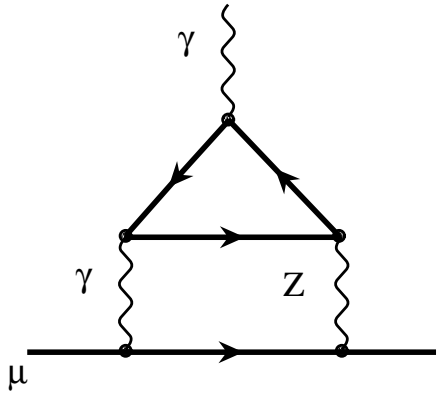}}
   \caption{\label{fig:fig2}
   \small Effective $Z\gamma\gamma^{*}$ coupling contributing to
   $a^{EW}_{\mu}$.}
\end{figure}

There is a class of electroweak contributions to the muon $g-2$
containing a fermion triangle along with a virtual photon and Z
boson as shown at Fig.~\ref{fig:fig2} that is discussed in
\cite{Kukhto, Peris, Knecht, Czarnecki1, Czarnecki2, Dorokhov}.
For the determination of the muon anomalous electromagnetic moment
we are interested in the $Z^{*}\rightarrow \gamma^{*}$ transition
in the presence of the external magnetic field to first order in
this field. In this approximation one may consider the current
$j_{\mu}$ as a source of a soft photon with polarization vector
$e^{\mu}(k)$ and momentum $k \rightarrow 0.$ In such kinematics
projection of the amplitude on $e^{\mu}(k)$ contains only two
Lorentz-invariant structures
\begin{eqnarray}
 T_{\alpha \nu}(p^{2},m^2) &=& T_{\alpha \mu \nu}(k,p) e^{\mu}(k)|_{k\rightarrow0}
 = w_{T}(p^{2},m^2)(-p^{2} \tilde{f}_{\nu \alpha} +
p_{\nu}p^{\rho} \tilde{f}_{\rho \alpha} -
     p_{\alpha}p^{\rho} \tilde{f}_{\rho \nu})+
\nonumber
\\
 &+& w_{L}(p^{2},m^2)p_{\alpha}p^{\rho}\tilde{f}_{\rho \nu},
\label{conf2}
\\
 \tilde{f}_{\mu \nu} &=& \frac12\varepsilon_{\mu \nu \gamma
     \delta} f^{\gamma \delta}, \qquad f_{\mu
     \nu}=k_{\mu}e_{\nu}-k_{\nu}e_{\mu}
 \nonumber
\end{eqnarray}
and can be viewed as a correlator of the axial and vector currents
in the external electromagnetic field with strength tensor $f_{\mu
\nu}.$ The same expression appears while analyzing the dominant
contribution of light-by-light scattering to $g-2$.

Both structures $w_{T}$ and $w_{L}$ are transversal with respect
to vector current, but only the first structure is transversal
with respect to axial current while the second is longitudinal.
According to the classical papers by Rosenberg \cite{Rosenberg}, Adler
\cite{Adler}, Bell and Jackiw \cite{BellJackiw} at the one-loop
level the invariant functions $w_{L, T}$ satisfy the relation
\begin{eqnarray}
 w_{L}^{(1-loop)}(p^2,m^2)=2w_{T}^{(1-loop)}(p^2,m^2).
 \label{rel1loop}
\end{eqnarray}
It was  shown by A.I. Vainshtein \cite{Vainshtein} that there is
the symmetry of the triangle amplitude $T_{\alpha \mu \nu}$ under
permutation $\nu \leftrightarrow \alpha$ in the chiral limit. This
symmetry preserves the relation between $w_{L}$ and $w_{T}$
(\ref{rel1loop}) for the case $m=0$ in any order of perturbation
theory.

Moreover, according to the Adler-Bardeen theorem \cite{AB} the
anomalous longitudinal part of the triangle is not renormalized in
the chiral limit.  It is worthy to note that this statement implies
an operator relation, while the matrix elements get the corrections \cite{AA} due to
anomalous dimension of axial current. At the same time, the
validity of Adler-Bardeen theorem at the operator level allows to
express \cite{ET} these three-loop corrections \cite{AA} in terms of earlier
two-loop and even one-loop calculations of anomalous divergencies.

To apply Adler-Bardeen theorem to the problem in question one should recall
that the axial
anomaly is expressed only through the longitudinal part $w_{L}$
\cite{Adler, BellJackiw}
\begin{eqnarray}
 p^{\alpha}T_{\alpha
 \nu}(p^2,m^2=0)=p^{2}w_{L}(p^2,m^2=0)\,p^{\sigma}\tilde{f}_{\sigma
 \nu}\sim p^{\sigma}\tilde{f}_{\sigma \nu}
 \label{anom}
\end{eqnarray}
and its nonrenormalization leads to the fact that the one-loop
result $w_{L}^{(1-loop)}\sim 1/p^{2}$ doesn't get the perturbative
corrections from gluon exchanges in the higher orders
\cite{Czarnecki3}. Nonrenormalization of $w_{L}$ implies the same
for $w_{T}$:
\begin{eqnarray}
w_{L}(p^2,m^2=0)=2w_{T}(p^2,m^2=0),\quad
w_{L,T}^{(>1-loop)}(p^2,m^2=0)=0. \label{theorem}
\end{eqnarray}
This nonrenormalization, in difference with longitudinal part,
holds on only perturbatively and, seemingly, is not directly
related to the phenomenon of anomaly. More general
nonrenormalization theorems in perturbative QCD were proved in
\cite{Knecht1}.

It is very interesting to look at the phenomenon of anomaly and
perturbative nonrenormalizability of $w_{L,T}$ using the
dispersive approach \cite{Zakharov1}. In framework of this
approach anomaly becomes quite simple and represents itself just
as an obvious subtraction constant. As a result,
the two-photon matrix element of the axial-vector current acquires
a pole in the chiral limit so the anomaly appears as a pure
infrared effect. Detailed investigation of the one-loop
VVA triangle graph within such approach has been performed earlier
\cite{Teryaev}.

The language of dispersion relations allows us to extract some new
specific properties of the higher order corrections to fermion
triangles. Relation (\ref{rel1loop}) in the context of dispersive
approach emerges due to the universality of anomaly when it is
appearing in the dispersion relation in the axial and vector
channels. This resembles the mentioned symmetry with respect to
the permutation $\nu \leftrightarrow \alpha$, observed by A.I.
Vainshtein. At the same time, here this relation is entirely
related to the anomaly phenomenon.

By direct analytical calculations of relevant formfactors in
two-loop approximation taking into account symmetry properties of
the amplitude we obtained that the full two-loop AVV amplitude is
equal to zero for arbitrary fermion masses. Non-anomalous Ward
identity for imaginary parts of formfactors has been proven that
provides the correctness of dispersive approach usage at least in
two-loop approximation. We also make a suggestion that the
dispersive approach is applicable in any order of perturbation
theory. Together with Adler-Bardeen theorem immediate consequence
of such suggestion is that the Vainshtein's theorem is correct for
nonzero fermion mass.

The paper is organized as follows. In Section 2 the dispersive
approach to the anomaly is briefly described for particular
configurations of the external momenta in the standard one-loop
triangle graph following to \cite{Teryaev}.  The Born approximation to
Vainshtein theorem in the framework of
dispersive approach is interpreted as equality of two expressions
for axial anomaly and valid in the case of finite fermion mass. In
section 3 the generalization of
dispersion approach to the axial anomaly for higher orders  of
perturbation is suggested.
The two-loop radiative corrections to anomalous triangle
with arbitrary fermion masses for the same kinematical
configurations are considered in the context of such an
approach. We found that all two-loop formfactors are zero,
justifying the postulated generalization of dispersive approach
and Vainshtein's theorem. Section 4 contains some concluding
remarks and discussion of the higher orders of perturbation theory and non-perturbative effects.
Appendix contains some details of the two-loop calculations.

\section{Axial anomaly and nonrenormalization theorem in the framework of dispersive approach}

\begin{figure}[h]
 \centerline{\includegraphics[width=0.3\textwidth]{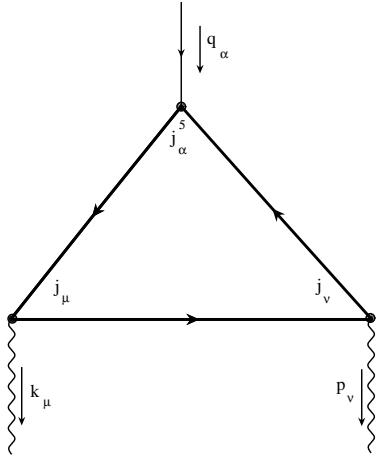}}
   \caption{\label{fig:fig1} \small Anomalous triangle diagram.}
\end{figure}

We use the standard tensor representation of the VVA triangle
graph amplitude (Fig.~\ref{fig:fig1}) due originally to Rosenberg
\cite{Rosenberg}
\begin{eqnarray}
    T_{\alpha \mu \nu}(k,p)
        &=& \varepsilon_{\alpha \mu \nu \rho}k^{\rho}F_{1}+\varepsilon_{\alpha \mu \nu \rho}p^{\rho}F_{2}
+ k_{\nu}\varepsilon_{\alpha \mu \rho
\sigma}k^{\rho}p^{\sigma}F_{3}+p_{\nu}\varepsilon_{\alpha \mu \rho
\sigma}k^{\rho}p^{\sigma}F_{4} \nonumber
\\
        &+&  k_{\mu}\varepsilon_{\alpha \nu \rho \sigma}k^{\rho}p^{\sigma}F_{5}+p_{\mu}\varepsilon_{\alpha \nu \rho
    \sigma}k^{\rho}p^{\sigma}F_{6}. \label{amp}
\end{eqnarray}

Here $F_{j}=F_{j}(q^{2};k^{2},p^{2},m^{2}),\;j=1,...,6$ are the
Lorentz invariant formfactors. The Bose symmetry of the amplitude
$T_{\alpha \mu \nu}(k,p)=T_{\alpha \nu \mu}(p,k)$ is equivalent to
\begin{eqnarray}
 F_{1}(k,p) &=&  -F_{2}(p,k) \nonumber
\\ \vspace*{5mm}
 F_{3}(k,p) &=&  -F_{6}(p,k)
\label{bose}
\\ \vspace*{5mm}
 F_{4}(k,p) &=&  -F_{5}(p,k). \nonumber
\end{eqnarray}

The gauge invariance leads to the vector Ward
identities $k^{\mu}T_{\alpha \mu \nu}=0,$ $p^{\nu}T_{\alpha \mu
\nu}=0$ which in terms of formfactors gives
\begin{eqnarray}
 F_{1} &=&  (kp) F_{3}+p^2 F_{4} \nonumber
\\ \vspace*{5mm}
 F_{2} &=&  k^{2} F_{5}+(kp) F_{6}.
\label{gauge}
\end{eqnarray}

In one-loop approximation all formfactors $F_{j}$ may be expressed
in Feynman-parametric form. For relevant formfactors we have
\begin{eqnarray}
 F_{3}(k,p)=-\frac{1}{\pi^2}I_{11}(k,p), \qquad
 F_{4}(k,p)=\frac{1}{\pi^2}[I_{20}(k,p)-I_{10}(k,p)],
 \label{feynpar}
\end{eqnarray}
where
\begin{eqnarray}
 I_{mn}(k,p)=\int^{1}_{0}dx\int^{1-x}_{0}dy\frac{x^{m}y^{n}}{y(1-y)k^{2}+
 x(1-x)p^{2}+2xy\,(kp)-m^{2}}
 \label{feynint}
\end{eqnarray}
Here $N_{c}=1$ for simplicity. These integrals have the
following symmetry properties
\begin{eqnarray}
 I_{mn}(k,p)=I_{nm}(p,k)
 \label{symint}
\end{eqnarray}
It is useful to observe that (\ref{bose}) together with (\ref{gauge})
and (\ref{symint}) implies
\begin{eqnarray}
 F_{6}(k,p)=-F_{3}(k,p)
 \label{symform}
\end{eqnarray}

It is well-known from the classic Adler's paper \cite{Adler} that
the accurate calculation of loop-momentum integrals at the
one-loop level leads to the anomalous axial-vector Ward identity
\begin{eqnarray}
 q^{\alpha}T_{\alpha \mu \nu}(k,p)=2mT_{\mu
 \nu}(k,p)+\frac{1}{2\pi^{2}}\varepsilon_{\mu \nu \rho
 \sigma}k^{\rho}p^{\sigma}.
 \label{anomWard}
\end{eqnarray}
The pseudotensor $T_{\mu \nu}$ may be written as
\begin{eqnarray}
 T_{\mu \nu}(k,p)=G(k,p)\varepsilon_{\mu \nu \rho \sigma}k^{\rho}p^{\sigma}.
 \label{axcur}
\end{eqnarray}
In terms of formfactors (\ref{anomWard}) reads
\begin{eqnarray}
 F_{2}-F_{1}=2mG+\frac{1}{2\pi^2}.
 \label{axcurform}
\end{eqnarray}

According to the Adler-Bardeen theorem the axial anomaly occurs
only at one-loop level. So, relation (\ref{axcurform}) for full
formfactors remains the same
\begin{eqnarray}
 F_{2}^{(tot)}-F_{1}^{(tot)}=2mG^{(tot)}+\frac{1}{2\pi^2}.
 \label{axcurformfull}
\end{eqnarray}

In the framework of dispersive approach we deal with the imaginary
parts of relevant formfactors.
We start with the kinematical configuration of
external momenta $k^{2}=0, \; p^{2}\ne 0$ for particular case
where $q^{2}>4m^2$ and $p^{2}<4m^2.$ Then there is a cut for
$q^{2}\in (4m^2,\infty)$ while there is no such singularity with
respect to variable $p^{2}.$ For the formfactors $F_{j}, \;
j=1,...,6$ and $G$ in any order of perturbation theory one may
write unsubtracted dispersion relations with respect to $q^{2}$
\begin{eqnarray}
 F_{j}(q^{2};p^{2},m^{2})
    &=& \frac{1}{\pi} \int_{4m^2}^{\infty} \frac{A^{A}_{j}(t;p^{2},m^{2})}{t-q^{2}}dt,
    \label{disrel}
\\ \vspace*{5mm}
 G(q^{2};p^{2},m^{2})
    &=& \frac{1}{\pi} \int_{4m^2}^{\infty} \frac{B^{A}(t;p^{2},m^{2})}{t-q^{2}}dt  \nonumber
\end{eqnarray}
We use the notations $F_{j}(q^{2};p^{2},m^{2})\equiv
F_{j}(q^{2};k^{2}=0,p^{2},m^{2})$ and $G(q^{2};p^{2},m^{2})\equiv
G(q^{2};k^{2}=0,p^{2},m^{2})$ below in the current section. The
$A^{A}_{j}$ and $B^{A}$ are the corresponding imaginary parts,
implying the cut with respect to variable $q^{2}=t,$ for example
\begin{eqnarray}
A^{A}_{j}(q^{2};p^{2},m^{2})=\frac{F_{j}(q^{2}+i\varepsilon;p^{2},m^{2})-F_{j}(q^{2}-i\varepsilon;p^{2},m^{2})}{2i}\nonumber
\end{eqnarray}


The imaginary parts of the relevant formfactors satisfy
non-anomalous Ward identities because they do not contain the
linear divergences in the momentum integrals
\begin{eqnarray}
 (p^{2}-t)A^A_{3}(t;p^{2},m^{2})-p^{2}A^A_{4}(t;p^{2},m^{2})=2mB^A(t;p^{2},m^{2})
 \label{nonanom}
\end{eqnarray}

Using (\ref{disrel}) and
(\ref{nonanom}) one gets finally
\begin{eqnarray}
  F_{2}(q^{2};p^{2},m^{2})-F_{1}(q^{2};p^{2},m^{2})-2mG(q^{2};p^{2},m^{2})=
\frac{1}{\pi}\int^{\infty}_{4m^2}A^A_{3}(t;p^{2},m^{2})dt
  \label{SR0}
\end{eqnarray}
Comparing with (\ref{axcurform}) and taking into account
(\ref{symform}) we find that the occurrence of the axial anomaly
at one-loop level is equivalent to a "sum rule" \cite{Teryaev}
\begin{eqnarray}
 \int^{\infty}_{4m^2}A_{3}^{A}(t;p^{2},m^{2})dt = \frac{1}{2\pi}
 \label{anomsumrule1}
\end{eqnarray}

It is easy to evaluate the $A_{3}^{A}$ by taking the imaginary
part of the corresponding integral in (\ref{feynpar}) \cite{Teryaev}
\begin{eqnarray}
 A_{3}^{A}(q^{2};p^{2},m^{2})=\frac{1}{2\pi}
 \frac{1}{(q^{2}-p^{2})^{2}}\left(-p^{2}R+2m^{2}\ln\frac{1+R}{1-R}\right),\;
 R=\left(1-\frac{4m^{2}}{q^{2}}\right)^{1/2}
 \label{A3}
\end{eqnarray}
By integration of this expression over $q^{2}$ one can check the
relation (\ref{anomsumrule1}).

In the preceding discussion we have employed dispersion relations
with respect to variable $q^{2}$ as these were appropriate for
considered kinematical region. Let us now discuss another version of
dispersive calculation of the anomaly using the cuts with respect to $p^2.$ For
this purpose we will consider another kinematical region
$p^{2}>4m^2, \; q^{2}<4m^2.$ Writing now instead of (\ref{disrel})
unsubtracted dispersion relations with respect to variable
$p^{2}=t$ for $F_{j}$ and $G$, we obtain
\begin{eqnarray}
  F_{2}(q^{2};p^{2},m^{2})-F_{1}(q^{2};p^{2},m^{2})-2mG(q^{2};p^{2},m^{2})=
\nonumber
\\
=\frac{1}{\pi}\int^{\infty}_{4m^2}
  [A^V_{4}(q^{2};t,m^{2})-A^V_{3}(q^{2};t,m^{2})]dt,
  \nonumber
\end{eqnarray}
The $A^{V}_{j}$ and $B^{V}$ are the corresponding imaginary parts,
implying the cut with respect to variable $p^{2}=t$
\begin{eqnarray}
A^{V}_{j}(q^{2};p^{2},m^{2})=\frac{F_{j}(q^{2};p^{2}+i\varepsilon,m^{2})-F_{j}(q^{2};p^{2}-i\varepsilon,m^{2})}{2i}.\nonumber
\end{eqnarray}
Thus, to recover the standard one-loop anomaly (\ref{axcurform})
taking into account (\ref{symform}) one has to show that
\begin{eqnarray}
 \int^{\infty}_{4m^2}[A_{4}^{V}(q^{2};t,m^{2})-A_{3}^{V}(q^{2};t,m^{2})]dt = \frac{1}{2\pi}
 \label{anomsumrule2}
\end{eqnarray}
for an arbitrary $m$ and for any considered value of $q^{2}.$

A straightforward calculation at one-loop level using
(\ref{feynpar}) gives a result \cite{Teryaev}
\begin{eqnarray}
 A_{3}^{V}(q^{2};p^{2},m^{2}) &=& \frac{1}{2\pi}
 \frac{1}{(p^{2}-q^{2})^{2}}\left(p^{2}S-2m^{2}\ln\frac{1+S}{1-S}\right),
\\ \vspace*{5mm}
 A_{4}^{V}(q^{2};p^{2},m^{2}) &=& \frac{1}{2\pi}
 \frac{1}{p^{2}-q^{2}}S,\quad
 S=\left(1-\frac{4m^{2}}{p^{2}}\right)^{1/2}
  \label{A3A4}
\end{eqnarray}
It was observed \cite{Teryaev} that the integrands
$A_{4}^{V}(q^{2};p^{2},m^{2})-A_{3}^{V}(q^{2};p^{2},m^{2})$
occurring in the sum rule (\ref{anomsumrule2}) at one loop level
are equal to the expression for $A_{3}^{A}(q^{2};p^{2},m^{2})$
from sum rule (\ref{anomsumrule1}) with $q^{2}$ and $p^{2}$ being
interchanged. As a result we have
\begin{eqnarray}
 A_{4}^{V}(q^{2};p^{2},m^{2})=A_{3}^{A}(p^{2};q^{2},m^{2})+A_{3}^{V}(q^{2};p^{2},m^{2}).
 \label{theorem2_1loop}
\end{eqnarray}

Let us write the unsubtracted dispersion relations with respect to
$q^2=t$ of the both sides of (\ref{theorem2_1loop}):
\begin{eqnarray}
 \frac{1}{\pi}\int^{\infty}_{4m^2} \frac{A^{V}_{4}(q^2;t,m^{2})}
 {t-p^{2}}dt=\frac{1}{\pi}\int^{\infty}_{4m^2}
 \frac{A^{A}_{3}(t;q^2,m^{2})}{t-p^{2}}dt+
\frac{1}{\pi}\int^{\infty}_{4m^2}
 \frac{A^{V}_{3}(q^2;t,m^{2})}{t-p^{2}}dt.
 \nonumber
\end{eqnarray}
One can immediately get from this expression that
\begin{eqnarray}
F_{4}(q^2;p^{2},m^{2})=F_{3}(p^{2};q^2;m^{2})+F_{3}(q^{2};p^2;m^{2}).
\label{theorem2F_1loop}
\end{eqnarray}

In the case with one soft $k\rightarrow 0$ and one virtual photons
$p^2\not=0$ we have $q^2=(k+p)^2\simeq p^2.$ For this kinematics
one can obtain the expressions for the longitudinal and
transversal parts of amplitude $T_{\alpha \nu}$ (\ref{conf2}) in
terms of Rosenberg's formfactors
\begin{eqnarray}
w_{L}(p^2,m^2)&=&F_{4}(p^2;p^2,m^2), \nonumber
\\
w_{T}(p^2,m^2)&=&F_{4}(p^2;p^2,m^2)-F_{3}(p^2;p^2,m^2).
 \label{omegaF}
\end{eqnarray}
The relation (\ref{rel1loop}) between $w_T$ and
$w_L$ in terms of formfactors
\begin{eqnarray}
F_{4}(p^2;p^{2},m^{2})=2F_{3}(p^2;p^2;m^{2}) \label{ARBJrelF}
\end{eqnarray}
immediately follows from (\ref{theorem2F_1loop}) with $q^2=p^2$.

In the framework of Vainshtein's approach the axial anomaly is expressed
only through the longitudinal part of triangle $w_{L}$ in the
chiral limit (\ref{anom}). Within the dispersive approach we have
two dispersive relations for axial anomaly (\ref{anomsumrule1}),
(\ref{anomsumrule2}) including imaginary parts of both structures
$w_{L}$ and $w_{T}$ for arbitrary mass.

\section{Calculation of two-loop axial anomaly and
check of dispersive approach and Vainshtein's theorem }

We propose the generalization of dispersion approach to the axial
anomaly for any order of perturbation theory. We suggest that
the imaginary parts of the relevant formfactors satisfy
non-anomalous Ward identities for higher order of perturbation
theory as well. This implies that anomaly will be also given by
corresponding finite subtractions.

\begin{figure}[h]
 \centerline{\includegraphics[width=0.7\textwidth]{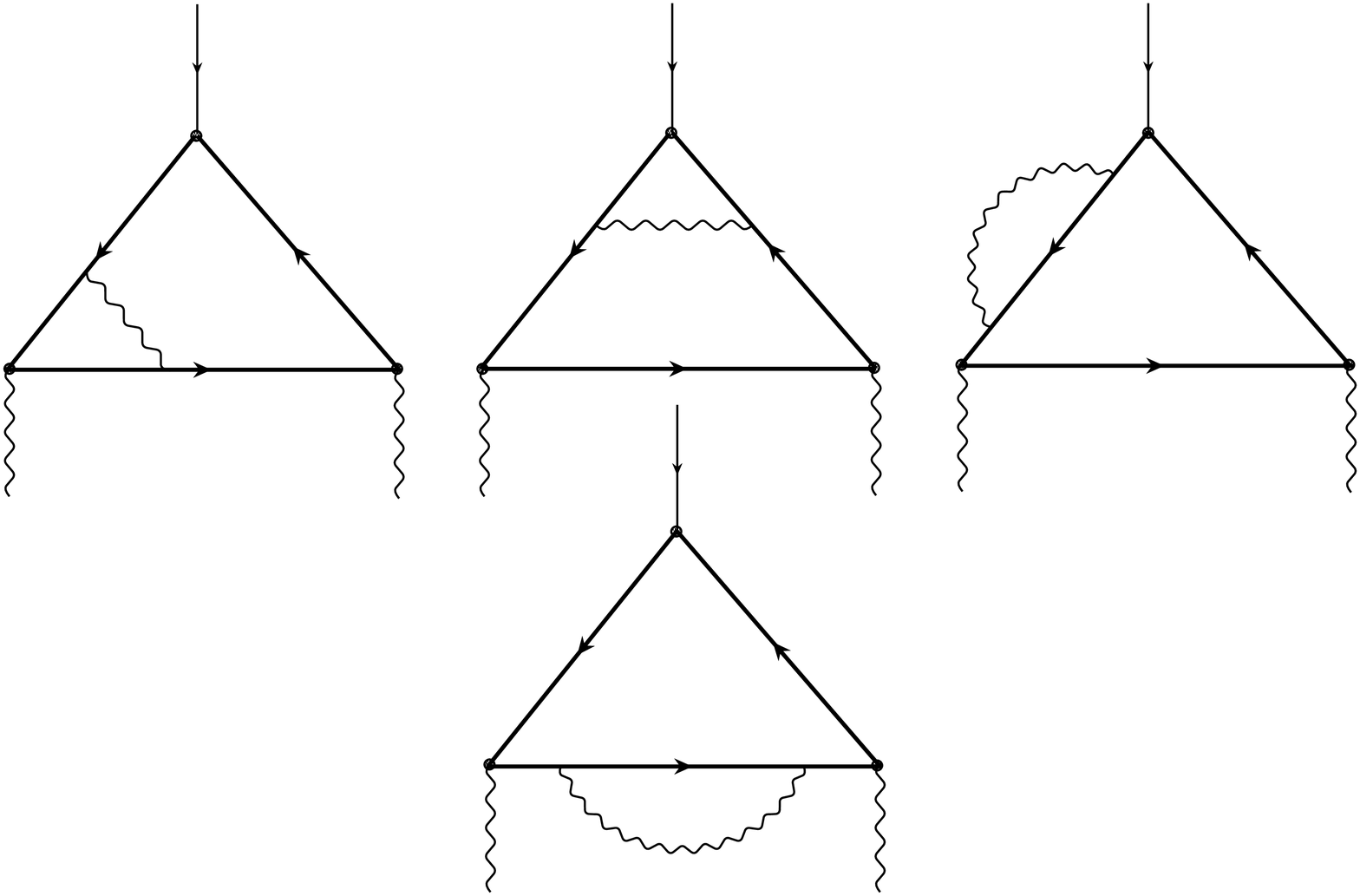}}
   \caption{\label{fig:fig3} \small Two-loop triangle diagrams.}
\end{figure}

To check that we will calculate the triangle diagram in two-loop approximation.
The results for QED and QCD corrections differ only by the obvious colour factor.
We consider the full amplitude of
anomalous triangles $T_{\alpha \mu \nu}^{(2-loop)}$ with all
possible types of radiative corrections shown at Fig.~\ref{fig:fig3}.

At first we construct four possible scalars from amplitude
$T_{\alpha \mu \nu}$ (\ref{amp}) in the particular kinematics
$k^{2}=0,\,p^{2}\not=0$:
\begin{eqnarray}
S_{1} &\equiv& T_{\alpha \mu
\nu}k^{\alpha}\varepsilon^{\xi\mu\nu\eta}k_{\xi}p_{\eta} = -
2(kp)^{2}F_{2}, \nonumber
\\
S_{2} &\equiv& T_{\alpha \mu
\nu}p^{\alpha}\varepsilon^{\xi\mu\nu\eta}k_{\xi}p_{\eta} =
2(kp)^{2}F_{1}, \nonumber
\\
S_{3} &\equiv& T_{\alpha \mu \nu}
\varepsilon^{\alpha\mu\nu\eta}k_{\eta} = 6(kp)F_{2} +
2(kp)^{2}F_{4}, \nonumber
\\
S_{4} &\equiv& T_{\alpha \mu \nu}
\varepsilon^{\alpha\mu\nu\eta}p_{\eta} = 6(kp)F_{1} +
2(kp)^{2}F_{5} - 2(kp)^{2}F_{3} + 6p^{2}F_{2}.
 \label{scalars}
\end{eqnarray}
Together with vector Ward identities for formfactors in the same
kinematics
\begin{eqnarray}
 F_{1} &=&  (kp) F_{3}+p^2 F_{4} \nonumber
\\ \vspace*{5mm}
 F_{2} &=&  (kp) F_{6}.
\label{gauge_part}
\end{eqnarray}
we have the closed system of equations for all formfactors
$F_{j}=F_{j}(q^{2};k^{2}=0,p^{2},m^{2}),\;j=1,...,6.$ Its solution
is
\begin{eqnarray}
F_1 = \frac{S_2}{2(kp)^2},\; F_2 = -\frac{S_1}{2(kp)^2},\; F_3 =
\frac{(kp)S_2-p^2(kp)S_3-3p^2S_1}{2(kp)^4},\nonumber
\\
F_4 = \frac{(kp)S_3+3S_1}{2(kp)^3},\; F_5 =
\frac{S_4(kp)-2S_2-p^2S_3}{2(kp)^3},\; F_6 = -\frac{S_1}{2(kp)^3}
\label{solut}
\end{eqnarray}

In the framework of dispersive approach we are interested in
calculation of imaginary parts of corresponding formfactors that
is $A_{j}^{A(2-loop)}(q^{2};k^{2}=0,p^{2},m^{2})$, so let's
calculate the imaginary parts of scalars
$S_{n}=S_{n}(q^{2};k^{2}=0,p^{2},m^{2}),\,n=1,...,4$, for example,
with respect to $q^{2}$
\begin{eqnarray}
\Delta_{n}^A(q^{2};k^{2}=0,p^{2},m^{2})=\frac1{2i}[S_{n}(q^{2}+i\varepsilon;k^{2}=0,p^{2},m^{2})
- S_{n}(q^{2}-i\varepsilon;k^{2}=0,p^{2},m^{2})]. \nonumber
\end{eqnarray}

We use the Pauli-Villars regularization with the parameter $M$.
After integration over loop momenta we have obtained the
expressions for
$\Delta_{n}^{A(2-loop)}(q^{2};k^{2}=0,p^{2},m^{2})$ in the form of
Feynman-parametric integrals (see Appendix). We drop all the terms
with the imaginary parts with respect to $q^2$ which do not
survive in the limit $M\rightarrow \infty.$

The next step is to calculate the imaginary part of
Feynman-parametric integrals and to get the expressions for
$\Delta_{n}^{A(2-loop)}(q^{2};k^{2}=0,p^{2},m^{2})$ in explicit
form before taking off the regularization. We get the imaginary
part of each integrand and then we integrate the
$\delta$-functions with complicated arguments analytically step by
step. After second or third integration all Feynman-parametric
integrals in $\Delta_{n}^{A(2-loop)}(q^{2};k^{2}=0,p^{2},m^{2})$
turned to zero, so
\begin{eqnarray}
\Delta_{n}^{A(2-loop)}(q^{2};k^{2}=0,p^{2},m^{2})\equiv 0,\quad
n=1,...,4. \label{Deltazeros}
\end{eqnarray}
According to (\ref{solut}) we see that the imaginary parts of all
formfactors with respect to $q^2$ turned to zero in the
considering kinematics
\begin{eqnarray}
A_{j}^{A(2-loop)}(q^{2};k^{2}=0,p^{2},m^{2})\equiv 0,\quad
j=1,...,6. \label{Azeros}
\end{eqnarray}
As a result the anomaly sum rule (\ref{anomsumrule1}) preserves
one-loop form also at two-loop level.

The investigation of the Vainshtein's theorem requires the
discontinuities of formfactors with the respect to variable $p^2$
(or $k^{2}$). Unsubtracted dispersion relations (\ref{disrel})
guarantee that all two-loop formfactors $F^{(2-loop)}_{j}$ are
equal to zero
\begin{eqnarray}
F^{(2-loop)}_{j}(q^{2};k^{2}=0,p^{2},m^{2})\equiv 0,\quad
j=1,...,6. \nonumber
\end{eqnarray}
So, the full two-loop AVV amplitude in the considering kinematics
is equal to zero. Consequently, the discontinuities with respect
to $p^2$ (or $k^{2}$) are also zero
\begin{eqnarray}
A^{V(2-loop)}_{j}(q^{2};k^{2}=0,p^{2},m^{2})\equiv 0,\quad
j=1,...,6, \nonumber
\end{eqnarray}
and the anomaly sum rule (\ref{anomsumrule2}) also preserves one-loop form.
This immediately leads to validity of Vainshtein's theorem with finite mass at two loops
\begin{eqnarray}
w_{L}(p^2,m^2)=2w_{T}(p^2,m^2),\quad
w_{L,T}^{(2-loop)}(p^2,m^2)=0. \label{ourtheorem}
\end{eqnarray}

For completeness we also explicitly checked the correctness of
non-anomalous Ward identities for imaginary parts. To do so we
have considered the imaginary part of pseudoscalar formfactor
$B^{A(2-loop)}(q^{2};k^{2},p^{2},m^{2})$ in the particular
kinematics. By direct diagrammatic calculations of the two-loop
amplitude $T_{\mu \nu}^{(2-loop)}$ we found that
\begin{eqnarray}
B^{A(2-loop)}(q^{2};k^{2}=0,p^{2},m^{2})\equiv0. \nonumber
\end{eqnarray}
So, taking into account (\ref{Azeros}), the relation
(\ref{nonanom}) does not obtain the perturbative corrections in
two-loop approximation.

\section{Discussion}

In our work the axial anomaly and Vainshtein's nonrenormalization
theorem are considered in the framework of dispersive approach. We
found that all two-loop contributions to formfactors
$F_{j}(q^{2};k^{2}=0,p^{2},m^{2})$ and
$G(q^{2};k^{2}=0,p^{2},m^{2})$ are equal to zero for arbitrary
fermion mass. It allows us to prove the suggested generalization
of dispersive approach to axial anomaly and to expand the
Vainshtein's nonrenormalization theorem for arbitrary fermion
masses in the triangle loop.

Although we thus proved these properties only at two-loop level,
they are likely to be valid at any order of perturbation theory.
Indeed, the validity of non-anomalous Ward identities for
imaginary parts is due to the absence of the linear divergencies
and should hold at all loops. This, together with Adler-Bardeen
theorem, would result in validity of anomaly sum rules. As soon as
the corrections to their integrands are zero in chiral limit due
to Vainshtein's theorem, it is rather hard to imagine the function
which is non-zero for finite mass case, while its integral is
still zero. In turn, all other functions should be also zero due
to gauge invariance.

At the same time, the further studies  of dispersive approach at
higher orders and (especially) beyond perturbative theory are of
most interest. One should note here the recent calculations in the framework of
instanton model \cite{Dorokhov2} leading to the exponential, rather than power
corrections to Vainshtein's theorem.

It seems that it is the nonlocality of this model, rather than
instanton specifics, that provides this exponential behavior. In
fact, it is analogous to the exponential falloff of transverse
momentum dependent parton distributions whose coordinate
description \cite{kt} bears similarity to the vacuum non-local
condensates. One should also recall another observation and
suggestion of \cite{Teryaev}, namely, that the local vacuum
condensates do not match the dispersive description of axial
anomaly and that the non-local condensates may improve the
situation.

We are grateful to A.E. Dorokhov, J. Ho\u{r}ej\u{s}\'i, M. Knecht,
J. Stern, O.V. Tarasov and A.I. Vainshtein for useful discussions
and comments. We are also indebted to anonymous Referee for the
valuable comments which allowed to eliminate the inconsistencies
in the first version of the proof which, however, did not affect
the final result. This work is partially supported by grant RFBR
03-02-16816.

{\it Note added}

After finishing of our work we learned about the preprint of
F. Jegerlehner and O. V. Tarasov  \cite{JT} whose conclusions are similar to ours
in the chiral limit for arbitrary external momenta.

\section{Appendix}

The imaginary parts of invariant scalars (\ref{scalars}) are given by the following Feynman-parametric
integrals

\begin{eqnarray*}
\lefteqn{\Delta_{1}^{A(2-loop)} \sim  Disc_{~q^2} \biggl[ M^4
\int\limits_{0}^{1} da \int\limits_{0}^{1-a} db
\int\limits_{0}^{1} dx \int\limits_{0}^{1-x} dy \; [ 4( 2(kp)X_1 -
p^2X_2 )( x + 2y - 1 ) ] + }
\\
& &{} + M^2 \int\limits_{0}^{1} da \int\limits_{0}^{1-a} db
\int\limits_{0}^{1} dx \int\limits_{0}^{1-x} dy \; [ ( 32y(kp)^2 +
16xa(kp)p^2 + 16b(kp)^2 - 32m^2y(kp) +
\\
& &{} + 8(kp)p^2 - 8m^2yp^2 - 16(kp)^2 + 16x(kp)^2 + 32ya(kp)p^2 -
8x(kp)p^2 - 32yb(kp)^2 -
\\
& &{} - 16xb(kp)^2 - 16m^2x(kp) - 16y(kp)p^2 - 16a(kp)p^2 )X_1 + (
4yp^4 - 8(kp)p^2 - 16yap^4 -
\\
& &{} - 8xap^4 - 8m^2x(kp) + 8ap^4 - 16m^2p^2 + 24m^2yp^2 +
16m^2xp^2 + 8y(kp)p^2 )X_2 + (4p^4 -
\\
& &{} - 8x(kp)p^2 - 4xp^4 - 8m^2p^2 - 4yp^4 + 8m^2yp^2 + 8m^2xp^2
+ 8(kp)p^2 - 8y(kp)p^2 )X_3 ] +
\\
& &{} +  \int\limits_{0}^{1} da \int\limits_{0}^{1-a} db
\int\limits_{0}^{1-a-b} dc \int\limits_{0}^{1} dx
\int\limits_{0}^{1-x} dy \; [ - 16(kp)^2m^2( 2m^2y - 2m^2 - p^2y +
p^2 + 2(kp) )X_4^2 +
\\
& &{} + 32m^4y(kp)^2X_5^2 ] +
\\
& &{} +  \int\limits_{0}^{1} da \int\limits_{0}^{1-a} db
\int\limits_{0}^{1} dx \int\limits_{0}^{1-x} dy \; [ 4m^2( -
8y(kp)^2 - 4xa(kp)p^2 - 4b(kp)^2 + 4m^2y(kp) - 2(kp)p^2 +
\\
& &{} + 2m^2yp^2 + 4(kp)^2 - 4x(kp)^2 - 8ya(kp)p^2 + 2x(kp)p^2 +
8yb(kp)^2 + 4xb(kp)^2 +
\\
& &{} + 2m^2x(kp) + 4y(kp)p^2 + 4a(kp)p^2 + 2m^2(kp) )X_1 + 4m^2(
- yp^4 + 2(kp)p^2 + 4yap^4 +
\\
& &{} + 2xap^4 + 2m^2x(kp) - 2ap^4 + 3m^2p^2 - 4m^2yp^2 - 3m^2xp^2
- 2y(kp)p^2 )X_2 + 4m^2( - p^4 +
\\
& &{} + 2x(kp)p^2 + xp^4 + 2m^2p^2 + yp^4 - 2m^2yp^2 - 2m^2xp^2 -
2(kp)p^2 + 2y(kp)p^2 )X_3 ] \biggr];
\end{eqnarray*}
\begin{eqnarray*}
\lefteqn{\Delta_{2}^{A(2-loop)} \sim  Disc_{~q^2} \biggl[ M^4
\int\limits_{0}^{1} da \int\limits_{0}^{1-a} db
\int\limits_{0}^{1} dx \int\limits_{0}^{1-x} dy \; [ - 4( 2(kp)X_1
- p^2X_2 )( x + 2y - 1 ) ] + }
\\
& &{} + M^2 \int\limits_{0}^{1} da \int\limits_{0}^{1-a} db
\int\limits_{0}^{1} dx \int\limits_{0}^{1-x} dy \; [ ( - 32y(kp)^2
+ 16(kp)^2 - 16xa(kp)p^2 - 8(kp)p^2 +
\\
& &{} + 16m^2y(kp) + 16m^2x(kp) + 8x(kp)p^2 - 16x(kp)^2 +
16xb(kp)^2 - 16b(kp)^2 + 16y(kp)p^2 +
\\
& &{} + 32yb(kp)^2 - 32ya(kp)p^2 - 8m^2p^2 - 16m^2(kp) +
16a(kp)p^2 )X_1 + ( - 24y(kp)p^2 +
\\
& &{} + 16yap^4 + 8(kp)p^2 - 8m^2xp^2 - 8ap^4 + 4xp^4 - 8m^2yp^2 -
12yp^4 + 8xap^4 )X_2 + ( - 4p^4 +
\\
& &{} + 4xp^4 + 8y(kp)p^2 - 8m^2yp^2 + 4yp^4 - 8m^2xp^2 + 8m^2p^2
+ 8x(kp)p^2 - 8(kp)p^2 )X_3 -
\\
& &{} - 8m^2x(kp)X_6 ] +
\\
& &{} +  \int\limits_{0}^{1} da \int\limits_{0}^{1-a} db
\int\limits_{0}^{1-a-b} dc \int\limits_{0}^{1} dx
\int\limits_{0}^{1-x} dy \; [ - 16(kp)^2m^2( - 2(kp)y - 2m^2 +
2(kp) + 2m^2y + p^2 +
\\
& &{} + xp^2)X_4^2 - 16(kp)^2m^2( 2m^2 - 4m^2y )X_5^2 ] +
\\
& &{} +  \int\limits_{0}^{1} da \int\limits_{0}^{1-a} db
\int\limits_{0}^{1} dx \int\limits_{0}^{1-x} dy \; [ - 4m^2(
4(kp)^2 - 4x(kp)^2 - 8y(kp)^2 + 4xb(kp)^2 - 2(kp)p^2 +
\\
& &{} + 2x(kp)p^2 + 8yb(kp)^2 - 4xa(kp)p^2 - 8ya(kp)p^2 +
4y(kp)p^2 + 4a(kp)p^2 + 2m^2x(kp) -
\\
& &{} - 2m^2p^2 - 2m^2(kp) - 4b(kp)^2 )X_1 - 4m^2( - 2ap^4 +
4yap^4 + 2(kp)p^2 - 6y(kp)p^2 -
\\
& &{} - m^2xp^2 - m^2p^2 + xp^4 - 3yp^4 + 2xap^4 )X_2 - 4m^2( -
p^4 + 2x(kp)p^2 + xp^4 + 2m^2p^2 +
\\
& &{} + yp^4 - 2m^2yp^2 - 2m^2xp^2 - 2(kp)p^2 + 2y(kp)p^2 )X_3 +
8m^4x(kp)X_6 ] \biggr];
\end{eqnarray*}
\begin{eqnarray*}
\lefteqn{\Delta_{3}^{A(2-loop)} \sim  Disc_{~q^2} \biggl[ M^4
\int\limits_{0}^{1} da \int\limits_{0}^{1-a} db
\int\limits_{0}^{1} dx \int\limits_{0}^{1-x} dy \; [ - 8( X_1 +
X_2 )( x + 2y - 1 ) ] + }
\\
& &{} + M^2 \int\limits_{0}^{1} da \int\limits_{0}^{1-a} db
\int\limits_{0}^{1} dx \int\limits_{0}^{1-x} dy \; [ ( - 32(kp)y +
32yb(kp) - 16xap^2 - 32yap^2 + 16p^2y -
\\
& &{} - 16x(kp) - 8p^2 + 16m^2x + 8xp^2 + 16(kp) + 32m^2y + 16m^2
- 16b(kp) + 16ap^2 +
\\
& &{} + 16xb(kp) )X_1 + ( - 32(kp) + 8xp^2 + 40p^2y + 16m^2x +
16m^2y - 16p^2 + 16x(kp) +
\\
& &{} + 16ap^2 + 48(kp)y - 32yap^2 - 16xap^2 )X_2 + ( - 8p^2y +
16m^2x + 16m^2y - 16(kp)y +
\\
& &{} + 8p^2 - 16m^2 + 16(kp))X_3 ] +
\\
& &{} +  \int\limits_{0}^{1} da \int\limits_{0}^{1-a} db
\int\limits_{0}^{1-a-b} dc \int\limits_{0}^{1} dx
\int\limits_{0}^{1-x} dy \; [ 16m^2(kp)( - 2(kp)y - 4m^2 + 4m^2y +
4(kp) + 2p^2 -
\\
& &{} - 2p^2y + xp^2 )X_4^2 - 64m^4y(kp)X_5^2 ] +
\\
& &{} +  \int\limits_{0}^{1} da \int\limits_{0}^{1-a} db
\int\limits_{0}^{1} dx \int\limits_{0}^{1-x} dy \; [ - 8m^2( -
2x(kp) - 4(kp)y + 4yb(kp) - 4yap^2 + 2p^2y - p^2 +
\\& &{} + m^2x + xp^2 + 2(kp) + 2m^2y + 3m^2 - 2b(kp) + 2ap^2 + 2xb(kp) -
2xap^2 )X_1 -
\\
& &{} - 8m^2( 5p^2y - 2p^2 + m^2x + xp^2 + m^2 - 4yap^2 - 4(kp) +
2ap^2 + 6(kp)y - 2xap^2 +
\\
& &{} + 2x(kp) )X_2 - 8m^2( 2m^2x - 2m^2 + 2m^2y - p^2y + 2(kp) -
2(kp)y + p^2)X_3 ] \biggr];
\end{eqnarray*}
\begin{eqnarray*}
\lefteqn{\Delta_{4}^{A(2-loop)} \sim  Disc_{~q^2} \biggl[ M^4
\int\limits_{0}^{1} da \int\limits_{0}^{1-a} db
\int\limits_{0}^{1} dx \int\limits_{0}^{1-x} dy \; [ 8( - 1 + 2y +
x )X_1 + 8( - 1 + y + x )X_6 ] + }
\\
& &{} + M^2 \int\limits_{0}^{1} da \int\limits_{0}^{1-a} db
\int\limits_{0}^{1} dx \int\limits_{0}^{1-x} dy \; [ ( 24p^2 +
16(kp) - 16x(kp) - 16m^2 + 16b(kp) - 24xp^2 -
\\
& &{} - 32m^2y - 32yb(kp) + 32yap^2 - 16ap^2 + 16xap^2 - 32p^2y -
16m^2x - 16xb(kp) )X_1 -
\\
& &{} - 16yp^2X_2 + (8p^2y - 16m^2x - 16(kp) + 16m^2 + 16(kp)y -
8p^2 - 16m^2y )X_3 +
\\
& &{} + ( - 16b(kp) - 16x(kp) + 16yb(kp) + 16xb(kp) - 16p^2y +
16p^2 + 32(kp) - 32(kp)y -
\\
& &{} - 16m^2x )X_6 ] +
\\
& &{} +  \int\limits_{0}^{1} da \int\limits_{0}^{1-a} db
\int\limits_{0}^{1-a-b} dc \int\limits_{0}^{1} dx
\int\limits_{0}^{1-x} dy \; [ 8( - 6m^2y(kp)p^2 + 8m^2(kp)p^2 +
4y(kp)p^4 -
\\
& &{} - 8m^4y(kp) - 2x(kp)p^4 - 8m^2(kp)^2 + 4y(kp)^2p^2 - p^6 +
2m^2xp^4 + 8m^2y(kp)^2 +
\\
& &{} + 8m^4(kp) + 8m^4yp^2 - 4(kp)p^4 - 4(kp)^2p^2 + yp^6 - xp^6
+ 6m^2p^4 - 4m^2x(kp)p^2 -
\\
& &{} - 8m^4p^2 - 6m^2yp^4 )X_4^2 + 8( - 4m^2y(kp)^2 + 2m^2(kp)p^2
+ 2m^2yp^4 - 8m^4(kp) -
\\& &{} - 8m^4yp^2 -
4m^2x(kp)p^2 - 4m^2x(kp)^2 + 16m^4y(kp) + 2m^2y(kp)p^2 +
4m^2(kp)^2 )X_5^2 ] +
\\
& &{} +  \int\limits_{0}^{1} da \int\limits_{0}^{1-a} db
\int\limits_{0}^{1} dx \int\limits_{0}^{1-x} dy \; [ 8m^2( - 2(kp)
+ 2x(kp) + 3m^2 - 2b(kp) - 3p^2 + 2m^2y +
\\
& &{} + 4yb(kp) - 4yap^2 + 2ap^2 - 2xap^2 + 4p^2y + m^2x + 2xb(kp)
+ 3xp^2 )X_1 + 16m^2yp^2X_2 +
\\
& &{} + 8m^2( 2m^2x - 2m^2 + 2m^2y - p^2y + 2(kp) - 2(kp)y + p^2
)X_3 + 8m^2( -2xb(kp) + 2b(kp) +
\\& &{}
+ 2p^2y + m^2 - m^2y + 2x(kp) + m^2x - 4(kp) - 2yb(kp) - 2p^2 +
4(kp)y )X_6 ] \biggr],
\end{eqnarray*}

where
\begin{eqnarray*}
& &{} X_1 = 1/( - m^2ya - m^2yb + m^2y - m^2 + 2ab(kp) + p^2a -
p^2a^2);
\\
& &{} X_2 = 1/( m^2xa + m^2xb - m^2x + m^2ya + m^2yb - m^2y - m^2a
- m^2b - 2xya(kp) -
\\
& &{} - 2xyb(kp) + 2xy(kp) - p^2xa - p^2xb + p^2x + p^2x^2a +
p^2x^2b - p^2x^2 + p^2a - p^2a^2);
\\
& &{} X_3 = 1/( m^2xa + m^2xb - m^2x + m^2ya + m^2yb - m^2y - m^2a
- m^2b - 2xya(kp) -
\\
& &{} - 2xyb(kp) + 2xy(kp) + 2ab(kp) - p^2xa - p^2xb + p^2x +
p^2x^2a + p^2x^2b - p^2x^2 +
\\
& &{} + p^2a - p^2a^2);
\\
& &{} X_4 = 1/( m^2xa + m^2xb + m^2xc - m^2x + m^2ya + m^2yb +
m^2yc - m^2y - m^2a - m^2b -
\\
& &{} - m^2c - 2xya(kp) - 2xyb(kp) - 2xyc(kp) + 2xy(kp) + 2ac(kp)
- p^2xa - p^2xb - p^2xc +
\\
& &{} + p^2x + p^2x^2a + p^2x^2b + p^2x^2c - p^2x^2 + p^2a -
p^2a^2);
\\
& &{} X_5 = 1/( - m^2ya - m^2yb - m^2yc + m^2y - m^2 + 2ac(kp) +
p^2a - p^2a^2);
\\
& &{} X_6 = 1/( 2xy(kp) - 2xyb(kp) - 2xya(kp) - m^2b - m^2a - m^2y
+ m^2yb + m^2ya - m^2x +
\\
& &{} + m^2xb + m^2xa + p^2x - p^2xb - p^2xa - p^2x^2 + p^2x^2b +
p^2x^2a).
\end{eqnarray*}


\begin{thebibliography}{100}
\bibitem{Kukhto}
  T.~V.~Kukhto, E.~A.~Kuraev, Z.~K.~Silagadze and A.~Schiller,
  Nucl.\ Phys.\ B {\bf 371}, 567 (1992).

\bibitem{Peris}
  S.~Peris, M.~Perrottet and E.~de Rafael,
  Phys.\ Lett.\ B {\bf 355}, 523 (1995)
  [arXiv:hep-ph/9505405].

\bibitem{Knecht}
  M.~Knecht, S.~Peris, M.~Perrottet and E.~De Rafael,
  JHEP {\bf 0211}, 003 (2002)
  [arXiv:hep-ph/0205102].

\bibitem{Czarnecki1}
  A.~Czarnecki, B.~Krause and W.~J.~Marciano,
  Phys.\ Rev.\ D {\bf 52}, 2619 (1995)

\bibitem{Czarnecki2}
  A.~Czarnecki, W.~J.~Marciano and A.~Vainshtein,
  Acta Phys.\ Polon.\ B {\bf 34}, 5669 (2003)

\bibitem{Dorokhov}
  A.~E.~Dorokhov,
  Phys.\ Rev.\ D {\bf 70}, 094011 (2004)
  [arXiv:hep-ph/0405153].

\bibitem{Rosenberg}
  L.~Rosenberg, Phys. Rev. \textbf{129}, 2786 (1963).

\bibitem{Adler}
  S.~L.~Adler,
  Phys.\ Rev.\  {\bf 177}, 2426 (1969).

\bibitem{BellJackiw}
  J.~S.~Bell and R.~Jackiw,
  Nuovo Cim.\ A {\bf 60}, 47 (1969).

\bibitem{Vainshtein}
  A.~Vainshtein,
  Phys.\ Lett.\ B {\bf 569}, 187 (2003)
  [arXiv:hep-ph/0212231].

\bibitem{AB}
  S.~L.~Adler and W.~A.~Bardeen,
  Phys.\ Rev.\  {\bf 182}, 1517 (1969); S.~L.~Adler,
  arXiv:hep-th/0405040.

\bibitem{AA} A.~A.~Anselm and A.~A.~Johansen,
  JETP Lett.\  {\bf 49}, 214 (1989).

\bibitem{ET}
  A.~V.~Efremov and O.~V.~Teryaev,
  Sov.\ J.\ Nucl.\ Phys.\  {\bf 51}, 943 (1990).

\bibitem{Czarnecki3}
  A.~Czarnecki, W.~J.~Marciano and A.~Vainshtein,
  Phys.\ Rev.\ D {\bf 67}, 073006 (2003)
  [arXiv:hep-ph/0212229].

\bibitem{Knecht1}
  M.~Knecht, S.~Peris, M.~Perrottet and E.~de Rafael,
  JHEP {\bf 0403}, 035 (2004)
  [arXiv:hep-ph/0311100].

\bibitem{Zakharov1}
  A.~D.~Dolgov and V.~I.~Zakharov,
  Nucl.\ Phys.\ B {\bf 27}, 525 (1971);
  V.~I.~Zakharov,
  Phys.\ Rev.\ D {\bf 42}, 1208 (1990).

\bibitem{Teryaev}
  J.~Ho\u{r}ej\u{s}\'i and O.~Teryaev,
  Z.\ Phys.\ C {\bf 65}, 691 (1995).

\bibitem{Dorokhov2}
  A.~E.~Dorokhov,
  JETP Lett.\  {\bf 82}, 1 (2005)
  [arXiv:hep-ph/0505196];
  Eur.\ Phys.\ J.\ C {\bf 42}, 309 (2005)
  [arXiv:hep-ph/0505007];
  Acta Phys. Polon. {\bf B 36} (2005) 
  [arXiv:hep-ph/0510297].

\bibitem{kt}
  O.~V.~Teryaev,
  Phys.\ Part.\ Nucl.\  {\bf 35}, S24 (2004).

\bibitem{JT}
  F.~Jegerlehner and O.~V.~Tarasov,
  arXiv:hep-ph/0510308.

\end{thebibliography}
\end{document}